# Microscopic anisotropy misestimation in spherical-mean single diffusion encoding MRI

Rafael Neto Henriques[1], Sune N Jespersen[2,3] and Noam Shemesh[1*]

[1]Champalimaud Neuroscience Programme, Champalimaud Centre for the Unknown, Lisbon, Portugal
[2]Center of Functionally Integrative Neuroscience (CFIN) and MINDLab, Clinical Institute, Aarhus University, Aarhus, Denmark.
[3]Department of Physics and Astronomy, Aarhus University, Aarhus, Denmark

**Abbreviations:** diffusion tensor imaging (DTI), fractional anisotropy (FA), microstructural fractional anisotropy (µFA), spherical mean technique (SMT), standard model (SM), single-pulsed diffusion encoding (SDE), double-pulsed diffusion encoding (DDE).

*****Corresponding author:** Dr. Noam Shemesh, Champalimaud Neuroscience Programme, Champalimaud Centre for the Unknown, Av. Brasilia 1400-038, Lisbon, Portugal. E-mail: noam.shemesh@neuro.fchampalimaud.org . Phone number: +351 210 480 000 ext. #4467.

# Abstract

**Purpose:** Microscopic fractional anisotropy (µFA) can disentangle microstructural information from orientation dispersion. While double diffusion encoding (DDE) MRI methods are widely used to extract accurate µFA, it has only recently been proposed that powder-averaged single diffusion encoding (SDE) signals, when coupled with the diffusion standard model (SM) and a set of constraints, could be used for µFA estimation. This study aims to evaluate µFA as derived from the spherical mean technique (SMT) set of constraints, as well as more generally for powder-averaged SM signals.

**Methods:** SDE experiments were performed at 16.4 T on an ex vivo mouse brain ($\Delta/\delta = 12/1.5$ ms). The µFA maps obtained from powder-averaged SDE signals were then compared to maps obtained from DDE-MRI experiments ($\Delta/\tau/\delta = 12/12/1.5$ ms), which allow a model-free estimation of µFA. Theory and simulations that consider different types of heterogeneity are presented for corroborating the experimental findings.

**Results:** µFA, as well as other estimates derived from powder-averaged SDE signals produced large deviations from the ground truth in both gray and white matter. Simulations revealed that these misestimations are likely a consequence of factors not considered by the underlying microstructural models (such as intercomponent and intracompartmental kurtosis).

**Conclusion:** Powder-averaged SMT and (2-component) SM are unable to accurately report µFA and other microstructural parameters in ex vivo tissues. Improper model assumptions and constraints can significantly compromise parameter specificity. Further developments and validations are required prior to implementation of these models in clinical or preclinical research.

# Introduction

Diffusion MRI is perhaps one of the most valuable reporters for dimensions much smaller than the MRI voxel size, and thus it has been extensively utilized to characterize microstructural changes both in health and in disease[1-5]. Numerous methods, mainly relying on the Stejskal-Tanner technique[6] – later dubbed more generally "single diffusion encoding" (SDE)[7] – were developed for characterizing diffusion anisotropy[8,9], leading to robust white matter (WM) orientation mapping in-vivo[5]. Other methods deriving diffusion anisotropy include, among others, quadratic-form analysis[10] or analysis of displacement profiles in q-space imaging (QSI)[11-13].

Diffusion anisotropy is highly dependent on the mesoscopic tissue organization[14,15]. An emerging diffusion MRI field of research strives to decouple microscopic diffusion properties from the mesoscopic tissue organization. Methods based on double diffusion encoding (DDE)[16] have been proposed for decoupling microscopic diffusion anisotropy (or its normalized form, microscopic fractional anisotropy (µFA)) from orientation dispersion using displacement correlations[17-29]. Isotropic diffusion encoding can also estimate µFA if multiple Gaussian components are assumed[30-36].

An alternative approach for decoupling microscopic diffusion properties from the mesoscopic tissue organization using the more commonly-available SDE methodology involves microstructural modelling[37-40]. This was pioneered by Stanisz et al. in the characterization of water diffusion in bovine optical nerve[41], and used early on to quantify directional uncertainty in tractography[42-44] and to estimate fiber caliber[45-47]. Typically, biophysical models assume that biological tissues can be represented by a sum of non-exchanging Gaussian diffusion components[48-53], and a specific instance of this class of models is the so-called "standard model" (SM)[40,54]. In this physical picture, the models can be expressed as a convolution between the mesoscopic orientation distribution function (ODF) and a kernel containing scalar microstructural parameters (e.g. neurite density, neurite diffusivity etc.)[49-56].

The SM fitting landscapes are rather flat[56,57], and therefore the number of scalar parameters that can be estimated can be limited, depending on the b-value regimes. For a small number of relatively low b-values, strong constraints must be imposed. For example, the Neurite Orientation Dispersion and Density Imaging (NODDI) model constrains the diffusion coefficients to constant values and assumes a pre-determined function (Watson[52] or Bingham[53] distributions) for the ODF. If higher b-values are reached, softer constraints are usually imposed: the Neurite Density Model (NDM) truncates spherical harmonics at order four to represent the ODF[49], while LEMONADE estimates all SM parameters using signal moments up to the 6th order[56].

It was realized early on that in powder-averaged systems[58-64], the ODF can be treated as a constant, and thus scalar microstructural parameters can be estimated independently to specific ODF constraints. Water diffusing in endosperm tissue[59], $^3$He gas diffusing in lungs[60], and metabolites diffusing (predominantly) in randomly oriented[61-64] neuronal dendrites and axons, were all assumed to constitute powder-averaged systems. More recently, it was noticed that, rather than using powder-averaged *systems*, one could use powder-averaged *signals* to remove the orientational complexity for estimating scalar signal parameters, such as µFA[28-36]. Recently, the signal powder-averaging approach was extended to SDE signals and coupled with a constraint of a single Gaussian component for estimating µFA in neural tissues[65]. The framework was later expanded to accommodate two (Gaussian) components representing intracellular and extracellular domains, such that the "axonal" volume fraction could be measured[66]. However, further constraints become necessary in such expansions (due to the relatively low and few b-values used): specifically, it was assumed that (1) extracellular diffusion follows the tortuosity model[67]; (2) axial diffusivities of intracellular and extracellular components are equal; and (3) intracellular radial diffusivity is zero (Table 1). The coupling between powder-averaged SDE signals and imposition of these specific constraints was termed Spherical Mean Technique (SMT)[65,66], and we refer to them hereafter as SMT1 and SMT2, respectively, where 1 or 2 represent the number of components assumed by each model. Some of these constraints can be released when higher b-values are reached[56].

Imposing constraints on microstructural models may be a practical necessity, but it may also corrupt the parameter specificity[40]. Indeed, the assumption that biological tissues can be well-described by a small number of Gaussian diffusion components might be overly simplistic, given the large heterogeneity of microscopic structures within a typical voxel containing axons, neurites, neuronal bodies, glia cell bodies, glial processes, myelin, etc.[1]. Moreover, even in relatively homogeneous voxels (e.g. voxels predominantly containing WM), diffusion heterogeneity might exist due to size polydispersity (e.g., distribution of axon diameter[68-71]) or nonzero cell body fractions[72]. In principle, such effects will lead to signal kurtosis, with two main physical origins: (1) for Gaussian diffusion, the diffusivity variance from multiple components can induce non-exponential signal decays[73-77], (intercomponent kurtosis); (2) restricted, time-dependent diffusion[78-86], which has been observed in biological tissues, can contradict the Gaussian assumption within individual microstructural compartments (intracompartmental kurtosis). Diffusion kurtosis can also arise from exchange[75], and any combination of the above-mentioned kurtosis factors can be envisioned in a given voxel.

Here, we aimed to investigate the validity and specificity of µFA derived from powder-averaged SDE signals. We present a few simple theoretical considerations of current SMT and powder-averaged SM assumptions and the associated parameter constraints vis-à-vis kurtosis.

Experiments in ex-vivo mouse brains show large discrepancies between DDE-derived µFA, and their SM-driven SDE counterparts. We explore in simulation how the violation of the above-mentioned constraints affect µFA estimation. Our findings suggest that powder-averaged SM estimates do not capture the ground-truth, and that intravoxel heterogeneity may be a culprit.

# Theory

**Definition of Microscopic Fractional Anisotropy**

Diffusion in biological tissues is often represented by an ensemble of microscopic components (or compartments)[30,40,77], with the $i^{th}$ component characterized by its own individual diffusion tensor $D_i$ (Figure 1). The ensemble-averaged diffusion tensor $D$ then equals the average of the individual diffusion tensors $D_i$ of all components contributing signal to the ensemble:

$$D = \langle D_i \rangle. \tag{1}$$

where brackets represent an ensemble average. The fractional anisotropy of $D$, (equivalent to DTI's FA) can be expressed as[31]:

$$FA = \sqrt{\frac{3}{2} \frac{V_\lambda(D)}{V_\lambda(D) + (Tr(D)/3)^2}} \tag{2}$$

where $Tr(D)$ is the trace of $D$, and $V_\lambda(D)$ is the variance of the eigenvalues of $D$. The µFA is the fractional anisotropy of each individual diffusion tensor $D_i$, given by[31,34]:

$$\mu FA_i = \sqrt{\frac{3}{2} \frac{V_\lambda(D_i)}{V_\lambda(D_i) + (Tr(D_i)/3)^2}} \ . \tag{3}$$

For a better understanding of the quantities defined by Equations 1-3, individual diffusion tensors $D_i$ and their respective ensemble diffusion tensor $D$ for five different environments are illustrated in Figure 1. The environment represented in Figures 1A and 1B consists of well-aligned structures with intracylindrical signal only, but with $\mu FA_i = FA = 0.80$ or $0.65$, respectively. Figure 1C shows that FA differs substantially from $\mu FA_i$ for finite orientation dispersion.

For the sake of simplicity, all microscopic diffusion tensors $D_i$ in Figure 1A-C are assumed to have constant $\mu FA_i$. A different and often used representation of white matter, consists of two diffusion tensors representing signals from intracylindrical and extracylindrical spaces, respectively (Figure 1D-E). In this case, it might be impractical to measure the microscopic fractional anisotropy $\mu FA_i$ of individual microscopic components. Therefore, it is useful to define the effective μFA as[31,34]:

$$\mu FA = \sqrt{\frac{3}{2} \frac{\langle V_\lambda(\boldsymbol{D}_i)\rangle}{\langle V_\lambda(\boldsymbol{D}_i)\rangle + (\langle Tr(\boldsymbol{D}_i)\rangle/3)^2}} \qquad (4)$$

where $\langle V_\lambda(\boldsymbol{D}_i)\rangle$ is the average eigenvalue variance of microscopic diffusion tensors $\boldsymbol{D}_i$ and $\langle Tr(\boldsymbol{D}_i)\rangle$ is the average trace of microscopic diffusion tensors $\boldsymbol{D}_i$. It is important to note that the effective μFA is not the simple average of $\mu FA_i$ (i.e. $\mu FA \neq \langle \mu FA_i\rangle$). Moreover, according to Equation 4, the effective μFA is expected to be invariant to tissue organization (as illustrated on Figure 1D,E). The definitions above are general for any environment containing an indefinite number of different types of diffusion components.

**μFA estimates from powder-averaged SDE signals**

In the original SMT work, the powder-averaged signal was constrained to one Gaussian component[65], i.e. it is assumed that any voxel can be fully described by one axial $\lambda_\parallel$ and one radial $\lambda_\perp$ diffusivity (Figure 1 A-C). Under this constraint (SMT1), the powder-averaged signal is given by[65]:

$$\bar{E}_{SMT1}(b) = e^{-b\lambda_\perp} \frac{\sqrt{\pi} \, \text{erf}(\sqrt{b(\lambda_\parallel - \lambda_\perp)})}{2\sqrt{b(\lambda_\parallel - \lambda_\perp)}} \qquad (5)$$

where $\bar{E}$ is the powder-averaged signal decay, $b$ is the b-value, and erf() is the error function. $\lambda_\parallel$ and $\lambda_\perp$ can then be obtained by fitting Equation 5 to $\bar{E}(b)$ measured with at least two non-zero b-values. From the estimated $\lambda_\parallel$ and $\lambda_\perp$ and under the same single-component assumption, μFA from SMT1 can be computed as[65]:

$$\mu FA^{SMT1} = \sqrt{\frac{(\lambda_\parallel - \lambda_\perp)^2}{(\lambda_\parallel^2 + 2\lambda_\perp^2)}}. \qquad (6)$$

Although Equation 6 is equivalent to Equations 2, 3 and 4 for systems containing exactly one component with constant $\lambda_\parallel$ and $\lambda_\perp$ values (Figure 1A-C), this is not the case for heterogeneous environments (e.g. Figure 1D-E). Equation 6 can be easily expanded to the standard model comprising two components (e.g., intracellular and extracellular domains). The powder-averaged signal for the 2-component standard model (SM) is

$$\bar{E}_{SM}(b) = fe^{-b\lambda_\perp^i} \frac{\sqrt{\pi} \operatorname{erf}\left(\sqrt{b(\lambda_\parallel^i-\lambda_\perp^i)}\right)}{2\sqrt{b(\lambda_\parallel^i-\lambda_\perp^i)}} + (1-f)e^{-b\lambda_\perp^e} \frac{\sqrt{\pi} \operatorname{erf}\left(\sqrt{b(\lambda_\parallel^e-\lambda_\perp^e)}\right)}{2\sqrt{b(\lambda_\parallel^e-\lambda_\perp^e)}} \tag{7}$$

where $\lambda_\parallel^i$ and $\lambda_\perp^i$ are the axial and radial diffusivities of the intracellular component, $\lambda_\parallel^e$ and $\lambda_\perp^e$ are the axial and radial diffusivities of the extracellular components and $f$ is volume fraction of the intracellular component. Introducing these model parameters into Equation 4, μFA for the 2-component SM can be calculated as:

$$\mu FA^{SM} = \sqrt{\frac{3f(\lambda_\parallel^i-\lambda_\perp^i)^2 + 3(1-f)(\lambda_\parallel^e-\lambda_\perp^e)^2}{2f(\lambda_\parallel^i-\lambda_\perp^i)^2 + 2(1-f)(\lambda_\parallel^e-\lambda_\perp^e)^2 + [f(\lambda_\parallel^i+2\lambda_\perp^i)+(1-f)(\lambda_\parallel^e+2\lambda_\perp^e)]^2}}. \tag{8}$$

As previously demonstrated[56,57], Equation 7 presents a flat fitting landscape. To increase the fit robustness, Kaden et al.[66] proposed to impose the following constraints: (1) tortuosity constraint[67], $\lambda_\perp^e = (1-f)\lambda_\parallel^e$; (2) equal axial diffusivities constraint, $\lambda_\parallel^i = \lambda_\parallel^e \equiv \lambda$; and (3) the stick constraint, $\lambda_\perp^i = 0$. We term this set of constraints SMT2 (Tab. 1). Equations 7 and 8 can be then rewritten as

$$\bar{E}_{SMT2}(b) = f \frac{\sqrt{\pi}}{2} \frac{\operatorname{erf}(\sqrt{b\lambda})}{\sqrt{b\lambda}} + (1-f)e^{-b(1-f)\lambda} \frac{\sqrt{\pi}}{2} \frac{\operatorname{erf}(\sqrt{bf\lambda})}{\sqrt{bf\lambda}}. \tag{9}$$

and

$$\mu FA^{SMT2} = \sqrt{\frac{3(1-2f_e^2+f_e^3)}{3+2f_e^2+4f_e^4}}. \tag{10}$$

where $f_e = (1-f)$.

Equation 10 shows that, under the constraints imposed by Kaden et al.[66], µFA only depends on the "neurite" fraction $f$. This theoretically reveals that Equation 10 cannot properly quantify µFA for arbitrary tissue configurations. For instance, for loosely packed randomly oriented fibers, the low extracellular diffusion anisotropy should naturally decrease the effective µFA. However, since $\mu FA^{SMT2}$ only depends on $f$, Equation 10 will fail to capture such effects.

**Cumulant expansion of the standard model powder-averaged signals**

To assess how kurtosis contributes to SM, it is useful to first express other easily-interpretable quantities under the assumption that SMT assumptions are valid. For instance, the second-order cumulant expansion[74,87] can be used to derive the effective diffusion $\bar{D}$ and effective kurtosis $\bar{K}$ of the powder-averaged signals:

$$\bar{E}(b) = \exp\left(-\bar{D}b + \frac{1}{6}\bar{K}\bar{D}^2 b^2 + O(b^3)\right) \quad (11)$$

For the one-component SM (SMT1), the theoretical diffusion $\bar{D}_{SMT1}$ and kurtosis $\bar{K}_{SMT1}$ of the powder-averaged signals can be calculated by expanding Equation 5 to second-order in $b$:

$$\bar{D}_{SMT1} = \frac{(\lambda_\| + 2\lambda_\perp)}{3} \quad (12)$$

and

$$\bar{K}_{SMT1} = \frac{4}{15} \frac{(\lambda_\| - \lambda_\perp)^2}{\bar{D}_{SMT1}^2} \quad (13)$$

Equation 13 shows that SMT1 only consider kurtosis from the variance across the eigenvalues of a single diffusion component.

Analogous to Equations 12 and 13, the diffusion $\bar{D}_{SM}$ and kurtosis $\bar{K}_{SM}$ for the two-component powder averaged SM can be computed by expanding Equation 7 to second order in $b$:

$$\bar{D}_{SM} = \frac{f\lambda_\|^i + (1-f)\lambda_\|^e + 2f\lambda_\perp^i + 2(1-f)\lambda_\perp^e}{3} \quad (14)$$

and

$$\overline{K}_{SM} = f\frac{4}{15}\frac{(\lambda_\parallel^i-\lambda_\perp^i)^2}{\overline{D}_{SM}^2} + (1-f)\frac{4}{15}\frac{(\lambda_\parallel^e-\lambda_\perp^e)^2}{\overline{D}_{SM}^2} + 3f(1-f)\frac{(\overline{D}^i-\overline{D}^e)^2}{\overline{D}_{SM}^2} \quad (15)$$

where the diffusivities $\overline{D}^i$ and $\overline{D}^e$ correspond to the mean diffusivities of the intracellular and extracellular components (i.e. $\overline{D}^i = (\lambda_\parallel^i + 2\lambda_\perp^i)/3$ and $\overline{D}^e = (\lambda_\parallel^e + 2\lambda_\perp^e)/3$. Note that the right-side terms of Equation 15 are related to the eigenvalues variances of both intracellular and extracellular components and the isotropic variance between $\overline{D}^i$ and $\overline{D}^e$.

Under SMT2 constraints, Equations 14 and 15 can be simplified to:

$$\overline{D}_{SMT2} = \frac{\lambda(1+2f_e^2)}{3} \quad (16)$$

and

$$\overline{K}_{SMT2} = \frac{216f-504f^2+504f^3-180f^4}{135-360f+420f^2-240f^3+60f^4} \quad (17)$$

Equation 17 shows that $\overline{K}_{SMT2}$ depends solely on $f$. Other kurtosis effects, e.g. arising from eigenvalue variance, are no longer accounted for.

## µFA estimates from powder-averaged DDE signals

In contrast to the SDE methodology, $\mu FA$ estimates from DDE ideally do not rely on specific microstructural model assumptions or parameter constraints. Jespersen et al. showed that the average eigenvalue variance $\langle V_\lambda(\boldsymbol{D}_i)\rangle$ can be computed from signals decays acquired for two parallel diffusion gradients $E_\parallel$ and signals decays acquired for two perpendicular diffusion gradients $E_\perp$[28]:

$$\log \overline{E}_\parallel/\overline{E}_\perp = \frac{3}{5}\langle V_\lambda(\boldsymbol{D}_i)\rangle b^2 + O(b^3) \quad (18)$$

where $\bar{E}_\parallel$ and $\bar{E}_\perp$ are the powder-averaged signals of $S_\parallel$ and $S_\perp$, using the 5-design orientation scheme[28]. The average eigenvalue variance extracted from Equation 18 can then be applied to Equation 4 to obtain a model-free estimate of μFA. In this study, $\langle Tr(\boldsymbol{D}_i) \rangle / 3$ of Equation 4 is set to the effective diffusion $\bar{D}$ computed from Equation 11.

# Methods

**Specimen preparation and MRI experiments**

All animal experiments were preapproved by the local ethics committee operating under local and EU law. A mouse brain was perfused intracardially from a healthy adult animal (N = 1), and was then immersed in 4% Paraformaldehyde (PFA) solution for 24 h, followed by Phosphate-buffered saline (PBS) solution for at least 48 h. The specimen was placed in a 10 mm NMR tube filled with Fluorinert (Sigma Aldrich, Lisbon, PT) and scanned at 37°C using a 16.4 T Bruker scanner (Karlsruhe, Germany) equipped with a Micro5 probe and a gradient system capable of producing up to 3000 mT/m in all directions.

For the generation of the gold-standard µFA map, diffusion-weighted signals were acquired using DDE-MRI for nineteen evenly sampled b-values from 0 to 4.5 ms/µm² ($\Delta$ = 12, $\delta$=1.5 ms, and mixing time = 12 ms) and 72 pairs of directions per b-value following Jespersen et al.'s 5-design[28]. To achieve a high SNR, acquisitions were repeated 14 times and averaged. Moreover, data for each b-value was further denoised using a Marchenko-Pastur-PCA denoising procedure (with an 8×8 sliding window)[88], while Gibbs ringing artefacts were suppressed using a sub-voxel shift algorithm[89]. µFA maps are then computed from the powder-averages of $E_\parallel$ and $E_\perp$ and using Equation 18 (higher-order corrections[90] were applied). It is important to note, that the DDE-MRI sequence was recently validated and shown to be robust against gradient artifacts (such as concomitant fields) under the experimental parameters used in this study[90].

The SDE experiments were then acquired for five evenly spaced b-values from 0.5 to 2.5 ms/µm² (2 averages), five b-values from 3 to 5 ms/µm² (4 averages), five b-values from 5.5 to 7.5 ms/µm² (10 averages), and three b-values from 8 to 9 ms/µm² (14 averages). The increasing number of averages were designed to compensate for the inherently lower SNR of high b-value data. For each b-value, SDE experiments were performed along 72 directions. Other MRI parameters common to both DDE and SDE experiments were as following: navigated 2-shot EPI acquisitions with double-sampling and a bandwidth of 384615 Hz; three coronal slices with thickness of 0.7 mm; FOV = 11×8.64 mm$^2$, Matrix Size = 92×72 leading to an in-plane resolution of 0.12×0.12 mm$^2$; partial Fourier factor = 1.25; TR/TE = 3000/50.4 ms.

The SDE data underwent the same preprocessing as in DDE (MPPCA and Gibbs unringing). The standard model was then used to extract µFA estimates from powder-averaged signals. For this, the four set of constraints (SMT1, SMT2, SM3, and SM4) summarized in Table 1 were tested. According to Novikov et al.[56] SM4 has two plausible branch solutions ($\lambda_\parallel^i > \lambda_\parallel^e$ and $\lambda_\parallel^i < \lambda_\parallel^e$) for low b-values, and we therefore fit data up to rather high b-values of 9 ms/µm$^2$. All models were fitted using grid search sampling parameters over 30 evenly-spaced values within their plausible ranges (diffusivities between 0 and 3ms/µm$^2$ and $f$ between 0 and 1). The set of parameters with lower sum of squared residuals are then used as the initial guess of a non-linear least squared fitting procedure which was implemented in Matlab® (The Mathworks, Nattick, USA).

**Simulations**

The effects of the model assumptions and parameter constraints were evaluated using five experiments based on synthetic data, where ground-truth µFA values are known a-priori. To assess the robustness of µFA estimates towards violation of model assumptions separately from the violation of parameter constraints, Simulations 1-3 were performed only for SMT1, which estimates both $\lambda_\parallel$ and $\lambda_\perp$ of a single component without imposing further constraints. The robustness of the two-component SM-based µFA estimates (SMT2, SM3 and SM4) were then assessed in Simulations 4 and 5. All simulations were generated noise-free. Synthetic signals were produced along the same gradient directions and for the same b-values as the ex-vivo SDE experiments. Each simulation is described in detail below.

   1. Multiple randomly oriented diffusion tensors (intercomponent kurtosis > 0): Intercomponent kurtosis was assessed by simulating an ensemble of 10000 randomly oriented microscopic components characterized by different tensors $D_i$ (no restricted diffusion, dispersion in the magnitude of $D_i$). The diffusion tensor of each microscopic component was generated by different combinations of both $\lambda_\parallel$ and $\lambda_\perp$ giving constant µFA$_i$ values (µFA$_{gt}$ = µFA$_i$) (gt, ground truth), but varying MD$_i$. The MD$_i$ values of the 10,000 components were randomly drawn from a Gaussian distribution with mean MD$_i$ of 0.8 µm$^2$/ms and a given MD$_i$ standard deviation (negative MD$_i$ values were zeroed). This experiment was also repeated for MD$_i$ values sampled from a bimodal Gaussian distribution with modes set to 0.8 and 1.7 µm$^2$/ms.

   2. Multiple diffusion tensors to represent a distribution of axons with non-zero radius (intercomponent kurtosis>0, no restricted diffusion): One way of treating white matter diffusion is to assume that diffusion transverse to axons of finite diameters is adequately described by the Gaussian approximation, hence producing a mapping between axon sizes and apparent transverse diffusivities. To assess kurtosis in this "intercomponent" case, diffusion-weighted signals of cylinders with different radii were generated using the MISST package[91]. Apparent transverse diffusivities were then set to the radial diffusivity obtained by fitting the standard DTI to individual MISST signals. Radii from a log-normal distribution with mean (*m*) and standard deviation (*std*) values were adjusted according to known diameters distributions in rat corpus callosum genu (*m* = 1 µm, *std* = 0.5 µm) and body (*m* = 1.5 µm, *std* = 0.7 µm), as well as for rat spinal cord white matter (*m* = 3.0 µm, *std* = 0.6 µm)[68,71]. The intrinsic diffusivity of these simulations was set to 2.5 µm$^2$/ms. Simulations were repeated for different cylinder lengths between 5 and 40 µm and repeated for an added extracellular Gaussian component with volume fraction $f_e = 0.3$, $\lambda_\parallel^e = 1.7$ μm²/ms, and $\lambda_\perp^e = 0.4$ μm²/ms. Ground-truth values were computed from a simulated DTI experiment on the same perfectly aligned cylinders.

3. <u>Multiple restricted cylindrical compartments to represent a distribution of axons with nonzero radius (intercompartmental and intracompartmental kurtosis >0):</u> Next, we simulated restricted diffusion in realistic compartment size distributions of white matter, containing both intercompartmental and intracompartmental kurtosis. For this, the signal of log-normally distributed cylinders was generated by directly averaging the signals obtained from the MISST package[91]. The dimensions of the cylinders and their log-normal distributions were identical to those described above[68,71]. This experiment was also repeated with an added extracellular Gaussian component.

4. <u>Testing the robustness of the SMT2 constraints:</u> To assess the robustness of SMT2 to the violation of its model constraints, synthetic signals are generated based on the two-component model, which represents extraneurite and intraneurite domains as two different diffusion tensors (no intracomponent kurtosis nor time-dependent diffusivities). The deviation of the tortuosity model was first tested by performing simulations for extraneurite $\lambda_\perp^e$ values, sampled between 0 and 2.0 μm²/ms independently of $f$, and while $\lambda_\parallel^i$ and $\lambda_\parallel^e$ were set to 2 μm²/ms. This experiment was then repeated for $\lambda_\parallel^i = 2.3$ μm²/ms and $\lambda_\parallel^e = 1.7$ μm²/ms (testing the violation of equal axial diffusivities).

5. <u>Multiple Gaussian components and restricted compartments to test the robustness of 2-component SM constraints:</u> The 2-component simulations were then applied to all three different sets of the two-component SM contraints (SMT2, SM3, SM4). To assess intercomponent kurtosis effects not considered by SM, synthetic signals were also generated for multiple components (as described in experiment 2) representing a distribution of axons with different radii together and an extracellular component with parameters set to $\lambda_\parallel^e = 1.7$ μm²/ms, $\lambda_\perp^e = 0.4$ μm²/ms, and $f_e = 0.3$. For the sake of simplicity, these simulations were performed only for log-normal radii distribution with *m*=1.5 μm and *std*=0.7μm. To assess the added effect of intracompartmental kurtosis, the latter simulation was then repeated by replacing the sum of Gaussian components with restricted diffusion cylindrical compartments (as described in experiment 3). Moreover, since previous studies suggested that brain tissues might also contain an isotropic restricted diffusion compartment (e.g. representing cell bodies)[41,47,92,93], the SM fits were further tested on a simulation where a fraction of isotropic compartments was added. For the sake of simplicity, the diffusivity of these isotropic compartment was set to zero (also known as a "dot" compartment)[93]. Ground-truth values were computed based on the apparent diffusivities of individual compartments computed from individual simulated DKI experiments[90].

To assess the robustness of the SM non-linear least squares fitting procedures to its initial guess, the simulations of experiment 5 were processed using a random initializer. This procedure was repeated for 100 sets of randomly sampled initial guess and compared to the grid search sampling fitting procedure.

# Results

**Experimental validation**

Figure 2A shows the µFA maps obtained for the three DDE data slices (hereafter referred to as $\mu FA^{DDE}$ maps). Under the assumption that DDE at the long mixing time regime provides a relatively model-free estimate of µFA, these maps can be considered as a "ground-truth" for this tissue. As expected, WM areas such as corpus callosum or internal capsule exhibit very high $\mu FA^{DDE}$, while GM areas exhibit lower $\mu FA^{DDE}$ (typically lower than 0.5). µFA maps calculated from SDE signals using the SMT1 constraints ($\mu FA^{SMT1}$) for the same slices are shown in Figure 2B. $\mu FA^{SMT1}$ values in the entire brain are much higher than $\mu FA^{DDE}$. Notice that $\mu FA^{SMT1}$ is still higher in WM compared with GM and approaches unity. Figure 2C similarly shows the µFA maps for SMT2 constraints ($\mu FA^{SMT2}$). In this case, $\mu FA^{SMT2}$ is clearly much lower than $\mu FA^{DDE}$ in GM, while $\mu FA^{SMT2}$ seems to be overestimated in WM. Figures 2D and 2E plot µFA from powder-averaged SDE fit using SM3 and SM4 constraints ($\mu FA^{SM3}$ and $\mu FA^{SM4}$, respectively). Both $\mu FA^{SM3}$ and $\mu FA^{SM4}$ appear to be overestimated compared to $\mu FA^{DDE}$ in GM and in WM. For reference, plots assessing the quality of the raw SDE and DDE data and standard DTI FA maps are shown in Supporting Information.

To quantify these aspects, Figure 3 displays µFA estimates for different sets of SM constraints plotted against the "ground-truth" $\mu FA^{DDE}$ from all voxels. For reference, identity and regression lines are marked by red dashed and black solid lines, respectively. For a better assessment of differences between WM and GM estimates, vertical lines that roughly segments these brain tissues estimates ($\mu FA^{DDE} = 0.5$) are shown. Compared to $\mu FA^{DDE}$, $\mu FA^{SMT1}$ appears to be overestimated always (Figure 3A), while $\mu FA^{SMT2}$ appears to be mainly underestimated, particularly for GM voxels (Figure 3B). $\mu FA^{SMT2}$ is also overestimated in the higher µFA regions (mainly in WM). $\mu FA^{SM3}$ and $\mu FA^{SM4}$ are closer to $\mu FA^{DDE}$ (Figure 3C and 3D, respectively); however, they are still higher than the gold-standard.

Parametric maps for all SMT2, SM3 and SM4 model parameters are shown in Figure 4. For all set of constraints, $f$ estimates are close to one in WM and close to zero in GM regions (Figure 4A1, B1, and C1 for SMT2, SM3 and SM4, respectively). $\lambda_\parallel$ maps derived from SM3 exhibit low contrast between gray and white matter regions (Figure 4B2), while $\lambda_\parallel^e$ maps calculated from SM4 (Figure 4C2) show lower values than $\lambda_\parallel^i$ maps (Figure 4C3). $\lambda_\perp^e$ estimated from SM3 (Figure 4B3) are similar to $\lambda_\perp^e$ extracted from SM4 (Figure 4C4). The relationship between different model parameter are assessed in the lower panels of Figure 4. In general, $\lambda_\perp^e$ extracted from both SM3 and SM4 are lower than the values predicted by the tortuosity model (Figure 4D1 and 4D2). No dependence was observed between $\lambda_\parallel^e$ and $\lambda_\parallel^i$ obtained from SM4 (Figure 4D3).

**Simulations**

1. Multiple randomly oriented diffusion tensors (intercomponent kurtosis > 0): Simulations for multiple randomly oriented diffusion tensor distributions are shown in Figure 5. Figure 5A1 shows unimodal distributions, as would be encountered for example in the intraneurite space alone. The corresponding estimates of $\lambda_\parallel^{SMT1}$, $\lambda_\perp^{SMT1}$, and $\mu FA^{SMT1}$ are plotted as a function of ground-truth values. Even for modest intercomponent kurtosis, $\lambda_\parallel^{SMT1}$ and $\lambda_\perp^{SMT1}$, are overestimated and underestimated, respectively, leading to a large overestimation of $\mu FA^{SMT1}$. To represent another distinct component, as would for example emerge from extracellular space, Figure 5B1 plots bimodal tensor distributions. Figure B2-4 show that the abovementioned trends for overestimates in $\lambda_\parallel^{SMT1}$, and underestimates in $\lambda_\perp^{SMT1}$ are much exacerbated, leading to a very large overestimation of $\mu FA^{SMT1}$.

2. Multiple diffusion tensors to represent a distribution of axons with non-zero radius (intercomponent kurtosis>0, no restricted diffusion): Figure 6 shows simulations for multiple Gaussian components, distributed under the assumption that axon sizes could be translated to Gaussian diffusivities. Simulated ground-truth $\lambda_\perp$ distributions from selected tissues are shown in Figure 6A. Figure 6B-C plots $\lambda_\parallel^{SMT1}$, $\lambda_\perp^{SMT1}$, and $\mu FA^{SMT1}$ as a function of the ground-truth $\lambda_\parallel$. When a hindered extracellular component is not considered (upper panel), $\lambda_\parallel^{SMT1}$, $\lambda_\perp^{SMT1}$, and $\mu FA^{SMT1}$ (solid lines) are quite close to their ground-truth values (dashed line, Figure 6B). However, as soon as another component is added, $\lambda_\parallel^{SMT1}$, $\lambda_\perp^{SMT1}$, and $\mu FA^{SMT1}$ are again subject to very large errors (Figure 6C).

3. Multiple cylindrical compartments to represent a distribution of axons with non-zero radius (intercompartmental and intracompartmental kurtosis >0): To investigate whether restricted diffusion – i.e. intracompartmental kurtosis – effects may be significant, Figure 7 presents simulated signals from realistic axon size distributions under the assumption of fully restricted diffusion within the axons. Figure 7A plots the distributions, which are identical to those used in Figure 6A. Figure 7 plots $\lambda_\parallel^{SMT1}$, $\lambda_\perp^{SMT1}$, and $\mu FA^{SMT1}$ (solid lines) and their ground-truth values (dashed lines), as a function of (finite) cylinder length. When only intracellular signals are considered, $\lambda_\perp^{SMT1}$ does not match ground-truth values (Figure 7B2), leading to underestimated $\mu FA^{SMT1}$ for cylinders with lengths smaller than 10 µm and overestimated $\mu FA^{SMT1}$ for larger lengths (Figure 7B3). When an extracellular component is added, $\lambda_\parallel^{SMT1}$ and $\lambda_\perp^{SMT1}$ are overestimated and underestimated, respectively (Figure 7C1 and 7C2). Consequently, $\mu FA^{SMT1}$ is overestimated independent of compartment length (Figure 7C3).

4. Robustness of the SMT2 parameter constraints: Figure 8A shows a perfect match between $\mu FA^{SMT2}$ (solid lines) and ground-truth values (dashed lines) for a two-component system produced exactly according to SMT2 constraints. However, as soon as the values of $\lambda_\perp^e$ and $f$ deviates from the tortuosity constraint ($\lambda_\perp^e = (1-f)\lambda_\parallel^e$), significant errors in $\lambda$, $f$ and $\mu FA^{SMT2}$ arise (Figure 8B). The red line signifies satisfied SMT2's tortuosity constraint, and $\lambda_\parallel^i = \lambda_\parallel^e = 2$ μm²/ms. Figure 8C shows the errors in $\lambda$, $f$ and $\mu FA^{SMT2}$ for arbitrary $\lambda_\perp^e$ and $f$ ground-truth values and for $\lambda_\parallel^i = 2.3$ μm²/ms and $\lambda_\parallel^e = 1.7$ μm²/ms. Notice the strong negative biases on $\lambda$, $f$ and $\mu FA^{SMT2}$ even when the tortuosity assumption is met (dashed red line).

5. Multiple Gaussian components and restricted compartments to test the robustness of two-component SM constraints: The scatter plots of Figure 9 show pairs of μFA and $f$ estimates obtained for different SM constraint sets and for different tissue scenarios. For each set of constraint (SMT2, SM3, SM4), estimates obtained from the grid search initialization approach are plotted by cross marked points, while estimates obtained from the random initialization are plotted by the dot marked points. For all tissue scenarios, $\mu FA^{SMT2}$ estimates from random initializations overlapped their respective estimate from the grid search approach. All $\mu FA^{SMT2}$ estimates are, however, lower than the ground truth value (black cross marked point in Figure 9A-D). The wide spread of $\mu FA^{SM3}$ and $\mu FA^{SM4}$ estimates obtained from the randomly initialized fitting approaches shows that SM3 and SM4 are associated to flat fitting landscapes. Nevertheless, $\mu FA^{SM3}$ and $\mu FA^{SM4}$ estimates obtained from the grid search approach are close to the μFA ground-truth value when intracellular and extracellular domains are represented by single diffusion tensors (Figure 9A). When both intercomponent kurtosis (Figure 9B) and intracompartmental kurtosis (Figure 9C) (e.g., arising from intracellular domains) are considered, biases are present in μFA/$f$ estimates of SM3 and SM4. These biases are dramatically higher when a fraction of zero-diffusion isotropic compartments is added (Figure 9D).

# Discussion

Quantifying diffusion properties independent of mesoscopic tissue organization is of high interest in many applications since it provides a more informative representation of underlying microstructural features[28-36,49-57]. µFA, typically obtainable from DDE acquisitions, quantifies diffusion anisotropy in the object's eigen-frame which is (ideally) not affected by potentially deleterious orientation dispersion effects[16-29]. Recently, spherically-averaged SM signals with different constraints have been proposed for quantifying µFA using the more commonly-available SDE-MRI without having to estimate the ODF[65-66]. This study's purpose was focused on assessing whether µFA could be accurately estimated using the recently proposed SMT frameworks[65-66] in an ex-vivo mouse brain. Additionally, we explored more generally whether 2-component SM powder-averaged fits could be appropriate for mapping µFA accurately.

In the long mixing time regime, DDE is widely established to accurately report on µFA without requiring prior assumptions[28-29]. Potentially confounding higher-order term effects[91] were accounted for in this study, making $µFA^{DDE}$ estimation more reliable. However, it should be noted that the long mixing time regime approximation needs to be fulfilled, and that exchange between different components is not accounted for. Still, $µFA^{DDE}$ can be (cautiously) considered as a "ground-truth" for the experimental study.

Our experiments clearly showed that, in a realistic tissue such as a mouse brain, $µFA^{SMT1}$ is crucially overestimated compared with $µFA^{DDE}$ (Figures 2 and 3). The source of this overestimation was revealed from the theory: it arises from any source of diffusion kurtosis (Equation 13), which "masquerades" as the non-mono-exponential decay underlying SMT1's parameter estimation. Our simulations fully corroborated these experimental findings and theoretical predictions showed that kurtosis leads to $\mu FA^{SMT1}$ overestimation. The exception is diffusion in very elongated structures with a large difference between $\lambda_\perp$ and $\lambda_\parallel$. Whether represented by Gaussian or non-Gaussian diffusion, $\mu FA^{SMT1}$ is rather accurate so long as "intra-cellular" diffusion is the only source of signal (c.f. Figures 6B and 7B); in this case, the deviations in $\lambda_\perp^{SMT1}$ estimates are not sufficiently large to skew $\mu FA^{SMT1}$. However, as soon as another component is introduced with a somewhat different diffusivity (e.g., "extracellular" diffusion), $\mu FA^{SMT1}$ becomes unreliable (Figures 6C and 7C). Hence, SMT1 can play a role in characterizing, e.g., diffusion of intracellular metabolites, as already done in[61,63], but not diffusion of ubiquitous water.

We then tested whether the more advanced framework incorporating (exactly) two components[66], termed SMT2, could provide better estimates of the underlying microstructure. Our experimental findings clearly showed that $\mu FA^{SMT2}$ was significantly underestimated compared with $\mu FA^{DDE}$ (Figures 2 and 3). From a theoretical perspective, this has been shown to originate from the direct relationship imposed by SMT2's tortuosity constraint: $\mu FA^{SMT2}$ exclusively depends on the stick volume fraction (Equation 10), unlike the $\mu FA^{SM}$ (Equation 8), which, as would be expected, depends also on the eigenvalues. If diffusivity is larger/smaller than that predicted by the tortuosity model, $\mu FA^{SMT2}$ can be either overestimated or underestimated, respectively (Figure 8B). Figure 4 suggests that in our experimental data (and in agreement with human in-vivo data[35,56,94]), the diffusivity is lower than that predicted by the tortuosity model. If other SMT2 assumptions are violated, such as fixing the axial diffusivities for both intra-"neurite" and extra-"neurite" (stick) domains to a single intrinsic diffusion coefficient $\lambda$, more than 50% error can be incurred (Figure 8C), in line with our experimental findings (Figures 2 and 3).

Another interesting question is whether powder-averaged signals subject to the commonly used 2-component SM fits could better correspond to the microstructural information reported by the model-free DDE. In the vast majority of our voxels, both SM3 and SM4 (acquired with b-values up to 9 ms/µm$^2$, which aids in avoiding branch selection and stabilizes the fitting landscape) overestimated µFA. Before making any conclusions, it is worth pointing out that we have used *powder-averaged* SM fits to keep consistent with SMT1 and SMT2 frameworks; however, such powder-averaged signals have an a-priori relatively flat fitting landscape, arising from the truncation of spherical harmonics at L=0[56], which may make the fit more difficult[56,57]. In the future, directional SM fits will be studied and reported, although it is clear that for powder-averaged systems[59-61], this approach will not work since the higher order terms are inherently zero. Nevertheless, assuming that the grid search of our powder-averaged SM fits was relatively effective (c.f. Figure 9A-C), our results likely suggest that the powder-averaged SM description – at least in ex-vivo tissues scanned under the experimental parameters in this study – may be incomplete for neural tissues. One reasonable explanation is that two components simply do not capture the tissue heterogeneity sufficiently well (Figure 9D). For example, cell bodies exist in both white and gray matter[72] and have nonzero volume fractions[92], and ex-vivo samples might present fully restricted diffusion (a non-zero fraction of "dot" compartments)[47,93]. These features could perhaps suggest that a third component may be required for models of diffusion in tissues. Another potential confound of SM is the stick assumption, which ignores axon size distributions in WM or dendrites and astrocytic branches in GM, whose radii can be far from zero.

One perhaps alarming property of fitting powder-averaged SDE signals, is that the maps they produce may seem very agreeable (Figure 2). Without independent validation, it would be difficult to infer that there is anything wrong with such maps. This underscores the importance of method-based vs. model-based comparisons. In fact, much can be learnt from the exercise of comparing such signals, including, perhaps, deciding on which biophysical model is correct for a given voxel; alternatively, DDE (or other method-driven) signals can be used to constrain biophysical model fits.

# Conclusions

Spherically-averaged SDE signals – whether strongly constrained by parameter fixing, such as SMT1 and SMT2, or more weakly constrained such as SM3 or SM4 – produce discrepant μFA compared to their DDE counterparts in neural tissue. The origin of these discrepancies is traced to the imposed constraints and in SMT1 and SMT2, mainly to signal kurtosis arising from any source of heterogeneity, such as distributions of tensors or restricted diffusion. The two-component, minimally-constrained powder-averaged SM framework also produces different estimates compared to $\mu FA^{DDE}$, suggesting that either the powder-averaged approach is not robust enough, or it is insufficient for describing the underlying microstructure sufficiently well. Metrics extracted from the powder-averaged standard models should be taken with great caution and validated based on data from histology or by further comparisons with model-free measures obtained from non-conventional diffusion acquisition schemes.

# Acknowledgements

This study was funded by the European Research Council (ERC) (agreement No. 679058). The authors acknowledge Dr. Daniel Nunes for the assistance in the preparation of the mouse's brain, and thank Dr. Jelle Veraart and Prof. Mark D. Does (Vanderbilt University) for insightful discussions and suggestions.

# Figures

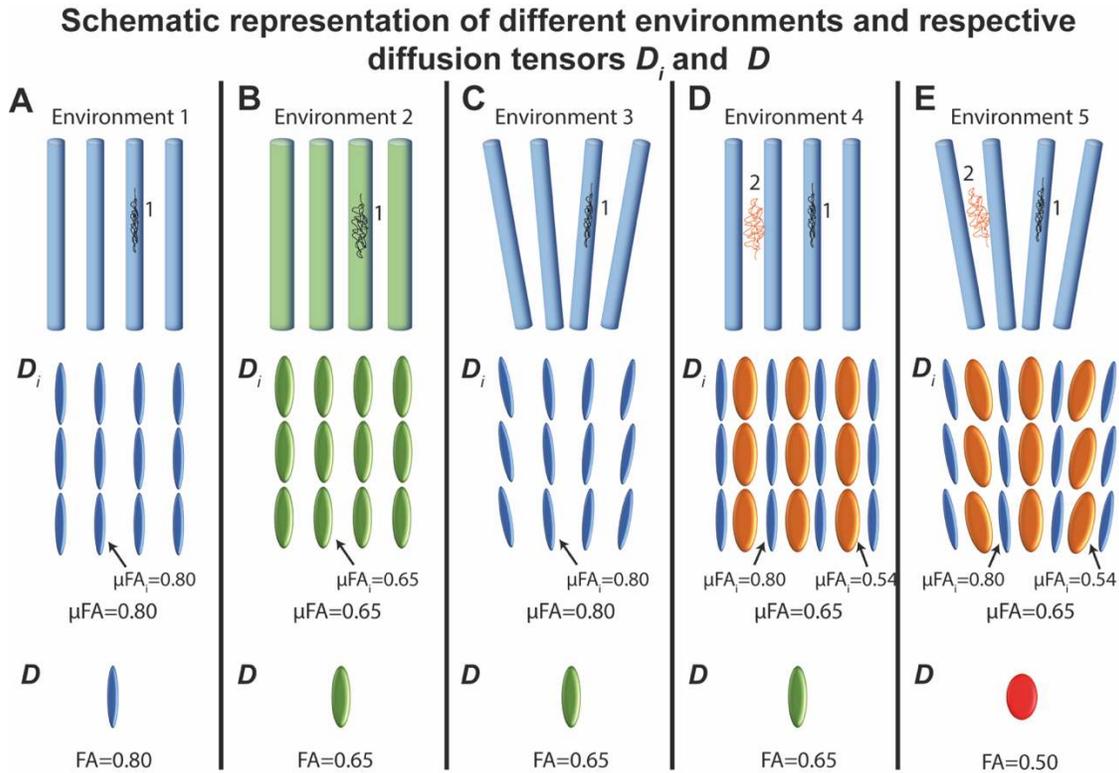

**Figure 1**. **Schematic representation of five different environments and respective individual diffusion tensors $D_i$ and ensemble diffusion tensor $D$**. All 5 microenvironments consist of cylindrical structures where water can diffuse in the interior and/or exterior spaces (random trajectories of water molecules diffusing in the interior and exterior of cylinders are illustrated by the lines marked by 1 and 2, respectively). **(A-C)** Considering only water inside the structures, diffusion might be represented by homogeneous individual tensor $D_i$ with constant microscopic fractional anisotropy ($\mu FA_i$). While the fractional anisotropy (FA) of tensor $D$ depends on both $\mu FA_i$ (differences between A and B) and dispersion orientation of the individual tensors $D_i$ (differences between A and C), the ensemble $\mu FA$ depends only on $\mu FA_i$. **(D and E)** Considering water inside and outside cylindrical microstructures, environments are better characterized by 2

types of tensors $D_i$. For these latter cases, µFA is still independent on orientation dispersion, and depends on the ensemble anisotropy of the diffusion tensors $D_i$.

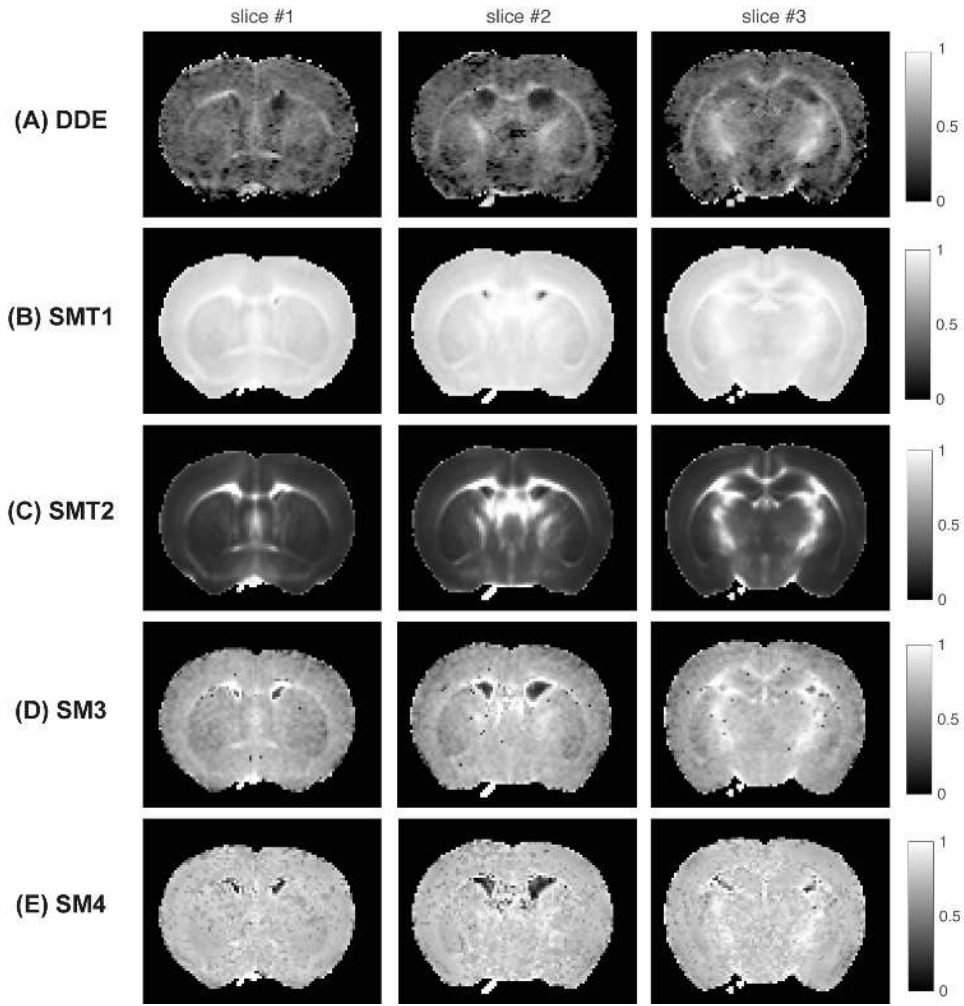

**Figure 2. Results for the ex-vivo mouse brain experiments: (A)** µFA gold-standard maps of the 3 DDE data slices; **(B)** µFA maps of the three SDE data obtained from SMT1; **(C)** µFA maps of the 3 SDE data slices obtained from SMT2; **(D)** µFA maps of the 3 SDE data slices obtained from SM3; **(E)** µFA maps of the 3 SDE data slices obtain from SM4.

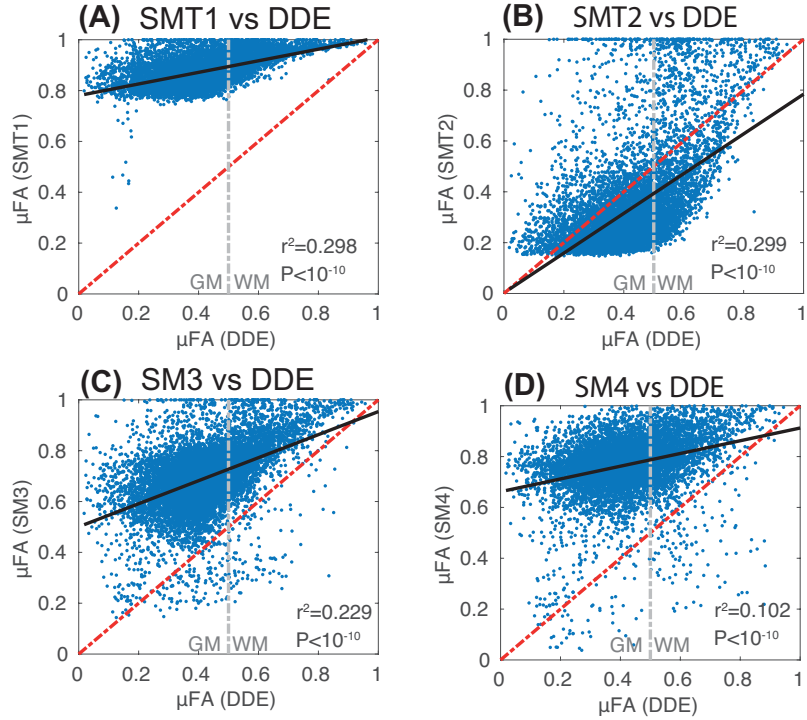

**Figure 3. Correlations between μFA estimates from SDE and DDE experiments: (A)** μFA estimates from SMT1 plotted as a function of the μFA estimates from DDE; **(B)** μFA estimates from SMT2 plotted as a function of the μFA estimates from DDE; **(C)** μFA estimates from SM3 plotted as a function of the μFA estimates from DDE; **(D)** μFA estimates from SM4 plotted as a function of the μFA estimates from DDE. For reference, the identity and regression lines are marked by the red dashed and black solid lines, while the DDE μ$FA$ value that roughly segments WM and GM are market by the gray vertical line.

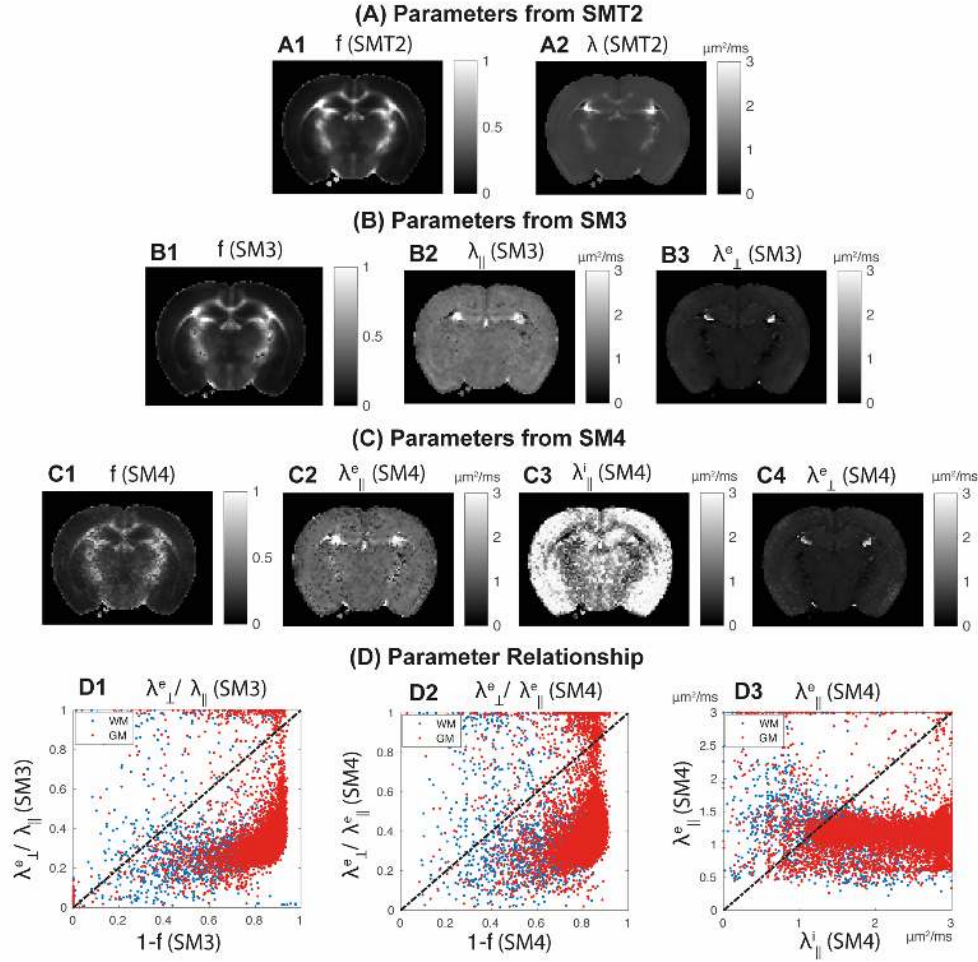

**Figure 4. SMT2, SM3 and SM4 parameters for the ex vivo mouse brain experiments. (A)** Parameter maps from SMT2: (A1) axonal volume fraction *f*; (A2) intrinsic diffusivity $\lambda$. **(B)** Parameter maps from SM3: (B1) axonal volume fraction *f*; (B2) axial diffusivity $\lambda_\|^e$; (B3) extracellular radial diffusivity $\lambda_\perp^e$. **(C)** Parameter maps from SM4: (C1) axonal volume fraction *f*, (C2) extracellular axial diffusivity $\lambda_\|^e$, extracellular axial diffusivity $\lambda_\|^i$, and (C4) extracellular radial diffusivity $\lambda_\perp^e$. **(D)** relationship across parameters: (D1) $\lambda_\perp^e/\lambda$ vs $1-f$ for SM3; (D2) $\lambda_\perp^e/\lambda_\|^e$ vs $1-f$ for SM4; and (D3) $\lambda_\|^i$ vs $\lambda_\|^e$ for SM4. In these latter plots, values corresponding to DDE µ$FA$ higher that 0.5 are plotted in blue (mainly WM voxels), while voxels with DDE µ$FA$ lower that 0.5 are plotted in red (mainly GM voxels).

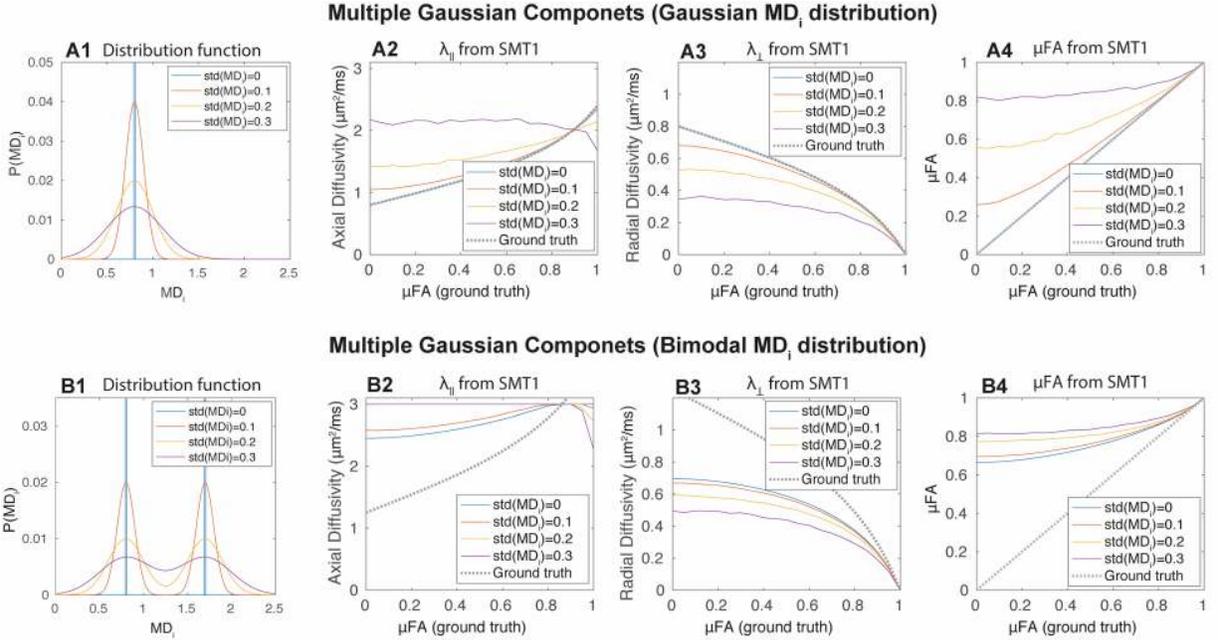

**Figure 5. Results from simulations generated based on multiple Gaussian componets: (A1)** Unimodal Gaussian distribution of the $MD_i$ values sampled for different standard deviations; **(A2)** SMT1 $\lambda_\parallel$ estimates for the unimodal distributions plotted as function of ground-truth µFA values; **(A3)** SMT1 $\lambda_\perp$ estimates for the unimodal distributions plotted as function of ground-truth µFA values; **(A4)** SMT1 µFA estimates for the unimodal distributions plotted as function of its ground-truth values. **(B1)** Bimodal Gaussian distribution of the $MD_i$ values sampled for different standard deviations; **(B2)** SMT1 $\lambda_\parallel$ estimates for the bimodal distributions from SMT1 plotted as function of ground-truth µFA values; **(B3)** SMT1 $\lambda_\perp$ estimates for the bimodal distributions from SMT1 plotted as function of µFA ground-truth values; **(B4)** SMT1 µFA estimates for the bimodal distributions from SMT1 plotted as function of its ground-truth values.

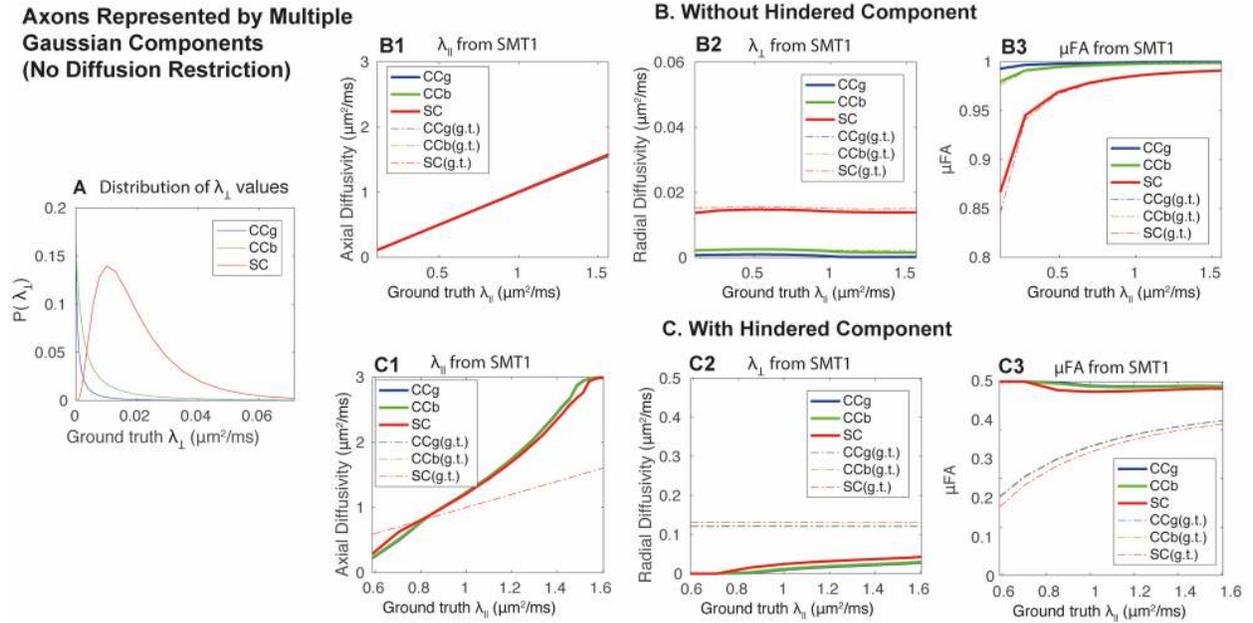

**Figure 6. Results from simulations generated based on multiple Gaussian components representing log-normal distributed axons:** **(A)** The 3 log-normal distributions of ground-truth radial diffusivities adjusted according to previous histological diameter measures of corpus callosum genu (CCg), corpus callosum body (CCb), and spinal cord (SC); **(B1)** $\lambda_\parallel$ estimates from SMT1 (solid lines) and $\lambda_\parallel$ ground-truth values (dashed lines) for the different diffusion distributions plotted as a function of $\lambda_\parallel$ ground-truth values; **(B2)** $\lambda_\perp$ estimates from SMT1 (solid lines) and $\lambda_\perp$ ground-truth values (dashed lines) for the different diffusion distributions and plotted as a function of $\lambda_\parallel$ ground-truth values; **(B3)** µFA estimates from SMT1 (solid lines) and µFA ground-truth values (dashed lines) for the different diffusion distributions and plotted as a function of $\lambda_\parallel$ ground-truth values. **(C1)** $\lambda_\parallel$ estimates from SMT1 (solid lines) and $\lambda_\parallel$ ground-truth values (dashed lines) for the different diffusion distributions involved by a single extracellular diffusion component; **(C2)** $\lambda_\perp$ estimates from SMT1 (solid lines) and $\lambda_\perp$ ground-truth values (dashed lines) for the different diffusion distributions involved by a single extracellular diffusion component;

**(C3)** μFA estimates from SMT1 (solid lines) and μFA ground-truth values (dashed lines) for the different diffusion distributions involved by a single extracellular diffusion component.

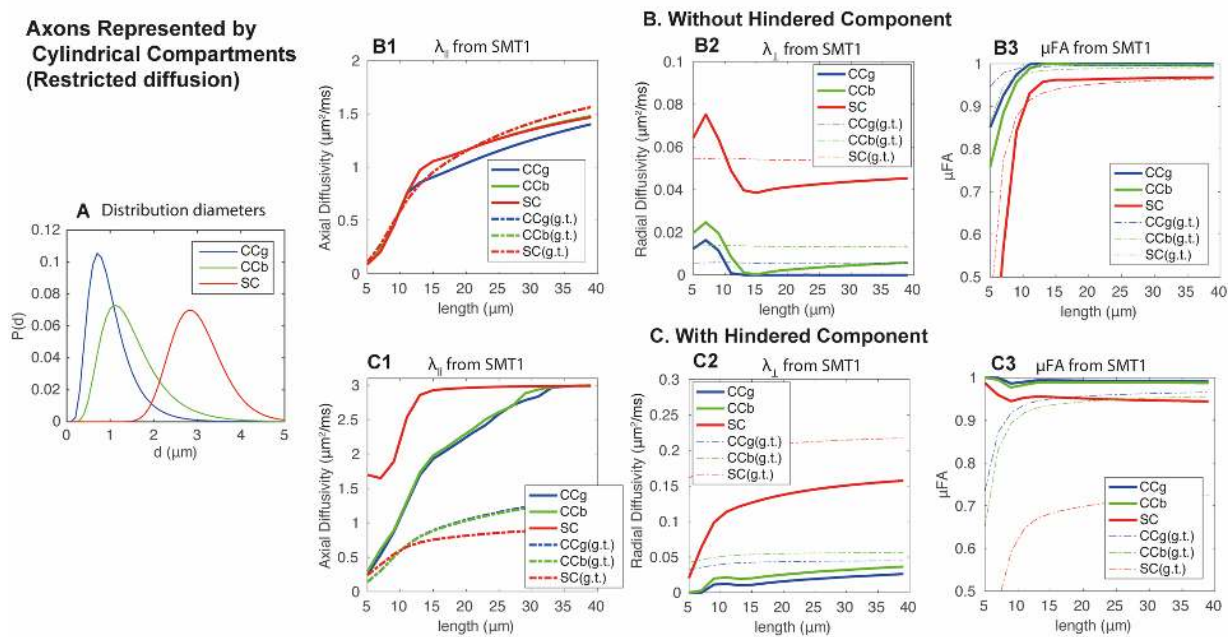

**Figure 7. Results from simulations generated based on multiple cylindrical compartments representing log-normal distributed axons:** **(A)** The 3 log-normal distributions of ground-truth diameters adjusted according to previous histological diameters measures of rat corpus callosum genu (CCg), rat corpus callosum body (CCb), and rat's spinal cord (SC); **(B1)** $\lambda_\parallel$ estimates from SMT1 (solid lines) and $\lambda_\parallel$ ground-truth values (dashed lines) for the different diameter distributions plotted as a function of ground-truth compartment length; **(B2)** $\lambda_\perp$ estimates from SMT1 (solid lines) and $\lambda_\perp$ ground-truth values (dashed lines) for the different diameter distributions and plotted as a function of ground-truth compartment length; **(B3)** μFA estimates from SMT1 (solid lines) and μFA ground-truth values (dashed lines) for the different diameter distributions and plotted as a function of ground truth compartment length. **(C1)** $\lambda_\parallel$ estimates from

SMT1 (solid lines) and $\lambda_\parallel$ ground-truth values (dashed lines) for the different diameter distributions involved by a single extracellular diffusion component; **(C2)** $\lambda_\perp$ estimates from SMT1 (solid lines) and $\lambda_\perp$ ground-truth values (dashed lines) for the different diameter distributions involved by a single extracellular diffusion component; **(B3)** µFA estimates from SMT1 (solid lines) and µFA ground-truth values (dashed lines) for the different diameter distributions involved by a single extracellular diffusion component.

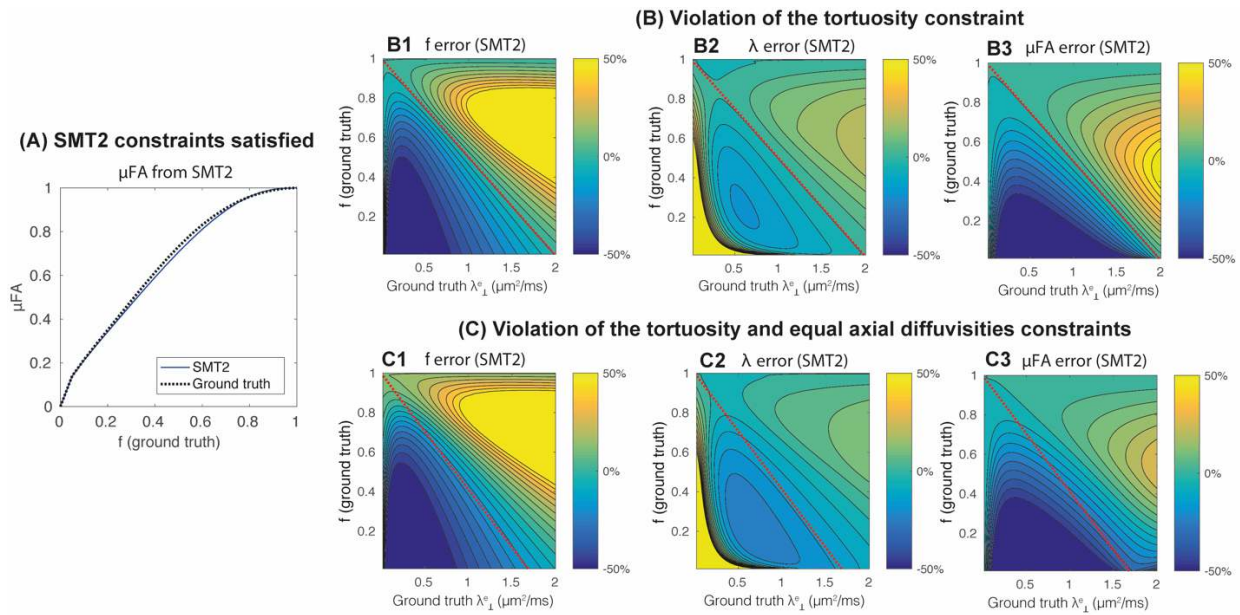

**Figure 8. Robustness of the SMT2 parameter constraints. (A)** µFA estimates from SMT2 when all SMT2 assumptions are met; **(B1)** Errors of the $f$ estimates from SMT2 when the tortuosity constraint is not met; **(B2)** Errors of the λ estimates from SMT2 when the tortuosity constraint is not met; **(B3)** Errors of the µFA estimates from SMT2 when the tortuosity constraint is not met. **(C1)** Errors of the $f$ estimates from SMT2 when the tortuosity and axial diffusivities constraints are not met; **(C2)** Errors of the λ estimates from SMT2 when the tortuosity and axial diffusivities constraints are not met; **(C3)** Errors of the µFA estimates from SMT2 when the tortuosity and axial

diffusivities constraints are not met. In panels B1-3 and C1-3, the ground-truth $\lambda_\perp^e$ and $f$ pair of values that match SMT2's tortuosity constraints are marked by the red dashed line.

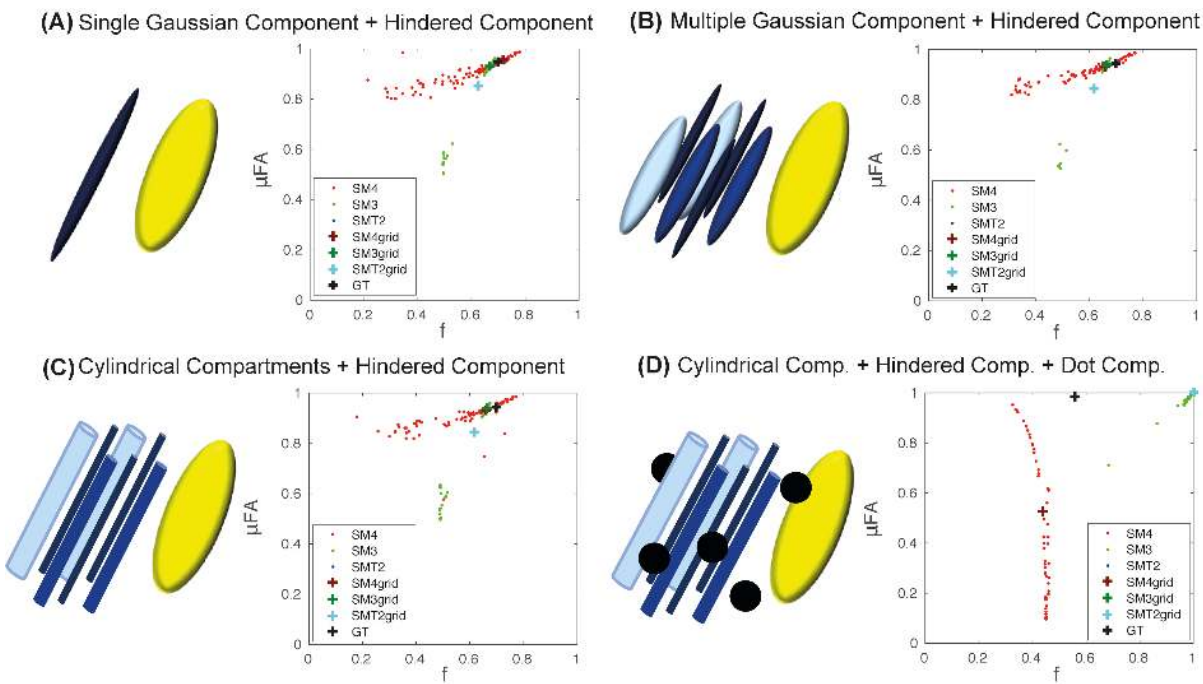

**Figure 9. Robustness of the standard model given different set of constraints (SMT1, SMT2, SM3, and SM4): (A)** Results for two Gaussian components representing intracellular ($\lambda_\parallel^i = 2.3$ μm²/ms, $\lambda_\perp^i = 0$ μm²/ms, $f = 0.7$) and extracellular ($\lambda_\parallel^e = 1.7$ μm²/ms, $\lambda_\perp^e = 0.4$ μm²/ms, $f_e = 0.3$) diffusivities; **(B)** Results for Gaussian components representing WM axons ($f = 0.7$ and with a log-normal radius distribution characterized by $m = 1.5$ μm and $std = 0.7$ μm) and a single component representing the extracellular ($\lambda_\parallel^e = 1.7$ μm²/ms, $\lambda_\perp^e = 0.4$ μm²/ms, $f_e = 0.3$) diffusivity; **(C)** Results for cylindrical compartments representing WM axons ($f = 0.7$ and with a log-normal radius distribution characterized by $m = 1.5$ μm and $std = 0.7$ μm) and a single component representing the extracellular ($\lambda_\parallel^e = 1.7$ μm²/ms, $\lambda_\perp^e = 0.4$ μm²/ms, $f_e = 0.3$) diffusivity. **(D)** Results for cylindrical compartments representing WM axons ($f = 0.56$ and with a log-normal radius distribution characterized by $m = 1.5$ μm and $std = 0.7$ μm), a single component representing the extracellular ($\lambda_\parallel^e = 1.7$ μm²/ms, $\lambda_\perp^e = 0.4$ μm²/ms, $f_e = 0.24$) diffusivity, and a fraction of zero-diffusion isotropic compartments ($f_r = 0.2$). For each panel, the pairs of μFA and $f$ values obtained from the grid search sampling fit procedure are plotted by cross markers "+"; the pairs of μFA and $f$ values obtained from parameters initialized randomly are plotted by the dot markers.

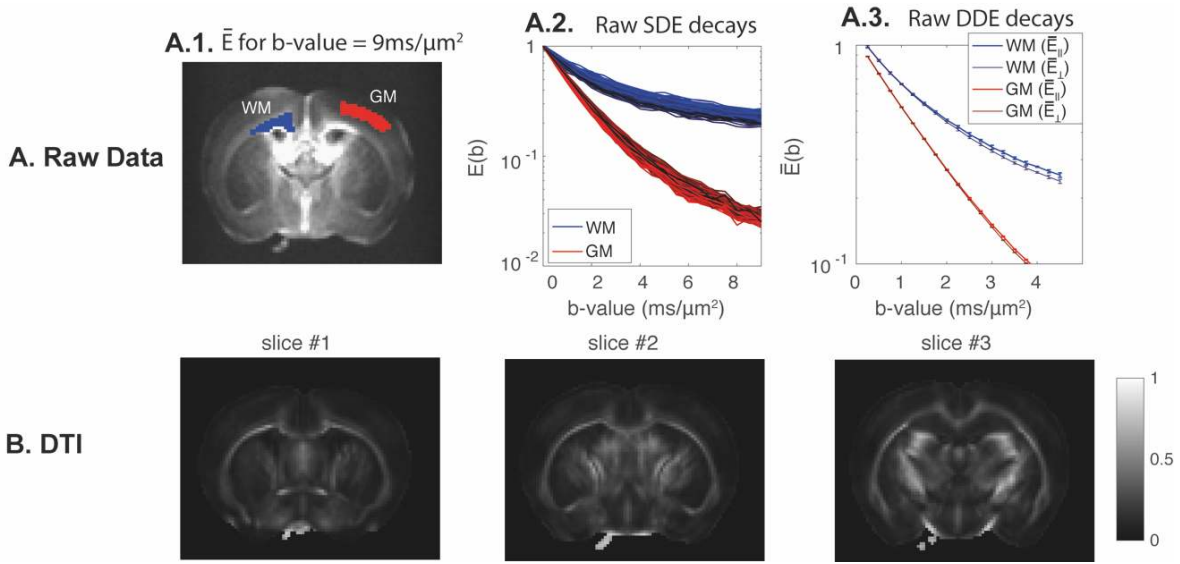

**Supporting Figure S1. The ex vivo mouse brain raw data and standard DTI FA maps: (A.1)** b-value = 0 map displaying the WM and GM regions used to plot the signal decays in panels A.2 and A.3. **(A.2)** raw signal decays for different SDE gradient directions and for WM and GM regions of interest **(A.3)** raw powder-averaged signal decays for parallel and perpendicular DDE gradient directions and for WM and GM regions of interest; **(B)** DTI FA maps of the three SDE data slices.

# Table

**Table 1. Summary of the four SM set of constraints used in this study.** Apart from the constraints mentioned, it is assumed that diffusion components have only 2 unique eigenvalues $\lambda_\parallel$ and $\lambda_\perp$.

| Name | Free model parameters | $n$ | Constraints |
|---|---|---|---|
| SMT1 | $\lambda_\parallel$ and $\lambda_\perp$ | 2 | 1) 1 component |
| SMT2 | $\lambda$ and $f$ | 2 | 1) 2 components <br> 2) $\lambda_\perp^e = (1-f)\lambda$ <br> 3) $\lambda_\parallel^i = \lambda_\parallel^e = \lambda$ <br> 4) $\lambda_\perp^i = 0$ |
| SM3 | $\lambda_\perp^e, \lambda$ and $f$ | 3 | 1) 2 components <br> 2) $\lambda_\parallel^i = \lambda_\parallel^e = \lambda$ <br> 3) $\lambda_\perp^i = 0$ |
| SM4 | $\lambda_\parallel^i, \lambda_\parallel^e, \lambda_\perp^e$ and $f$ | 4 | 1) 2 components <br> 2) $\lambda_\perp^i = 0$ |